\documentclass[a4paper,11pt]{article}
\usepackage{pos}

\usepackage{graphicx}
\usepackage{epsfig}
\usepackage{float}
\usepackage{braket}
\usepackage{slashed}
\usepackage{eqnarray,amsmath}

\title{Full NNLO QCD corrections to diphoton production}

\author*[a]{Federico Coro}

\affiliation[a]{Instituto de Fisica Corpuscolar, Universitat de Valencia - Consejo Superior de Investigaciones Cientificas,\\
 Parc Cientific, E-46980 Paterna, Valencia, Spain}

\emailAdd{fcoro@ific.uv.es}

\abstract{We consider the diphoton production in hadronic collisions at the next-to-next-to-leading order (NNLO) in perturbative QCD, taking into account for the first time the full top quark mass dependence. We present the computation of the two-loop form factors for diphoton production in the quark annihilation channel, that are relevant for the phenomenological studies of the full NNLO. The MIs are written in the so-called canonical logarithmic form, except for the elliptic ones. We perform a study on the Maximal Cut in order to show the elliptic behaviour of the non-planar topology. The Master Integrals are evaluated by means of differential equations in a semy-analitical approach through the generalised power series technique. Finally we use this result with all the other contributions showing selected numerical distributions. }

\FullConference{The European Physical Society Conference on High Energy Physics (EPS-HEP2023)\\
 21-25 August 2023\\
Hamburg, Germany\\}

\begin{document}
\maketitle

\section{Introduction}
\noindent The production of two prompt photons is one of the most important processes at the LHC for several reasons. Due to it’s cleanliness, it constitute an irreducible background for the signal made by the Higgs, and in particular it played a crucial role in the discovery of the Higgs itself. For the same reason, diphoton production is important to misure the fundamental parameters within the Standard Model and at the same time to search for new physics. 
Prior to our work, the state of the art for this process included scattering amplitudes in the massless case up to NNLO \cite{Catani:2018krb,Catani:2011qz,Campbell:2016yrh,Schuermann:2022qdm} and for the massive case the $gg \to \gamma \gamma$ with partial N$^{3}$LO contributions \cite{Bern:2001df,Maltoni:2018zvp,Bern:2002jx,Chen:2019fla}. Furthermore, amplitudes for $\gamma \gamma + jets$ \cite{ Chawdhry:2021mkw,Agarwal:2021vdh,Chawdhry:2021hkp,Badger:2021ohm,Badger:2021ega} and form factors up to 3 loops for massless loops \cite{Caola:2020dfu} have been known in literature. Considering the full NNLO accuracy, there were some missing elements. 
We discuss the calculation of our main contribution, i.e. the two-loop massive form factors for the partonic process $q \overline{q} \to \gamma \gamma$ \cite{Becchetti:2023wev} and then we use this result for our phenomenological purpose. We consider for the first time diphoton production at NNLO, taking into account the full top quark mass dependence \cite{Becchetti:ta}, including the Double-Virtual, Double-Real and Real-Virtual contributions.

\section{Two-Loop Form Factors}

\noindent In this section we consider the two-loop form factors for diphoton production in the quark annihilation channel with a heavy quark loop. 
At the partonic level, the scattering amplitude is given by:
\begin{equation}
q(p_1) + \overline{q}(p_2) \to \gamma(p_3) + \gamma(p_4).
\end{equation}

\noindent Our kinematics, for the $2 \to 2$ process, is described by the following Mandelstam invariants:
\begin{equation} \label{eq:mandelstam}
s=-(p_1+p_2)^2, \; t=-(p_1-p_3)^2, \; u=-(p_2-p_3)^2, \; \text{with} \; s+t+u = 0,
\end{equation}
where the external particles are on-shell, i.e. $p_{i}^{2}=0$.

\noindent Using projectors method \cite{Peraro:2020sfm}, at any order in QCD perturbation theory, the amplitude can be decomposed as :
\begin{equation}
\mathcal{A}_{q \overline{q}, \gamma \gamma}(s,t,m_t^{2}) = \sum_{i=1}^{4} \mathcal{F}_{i}(s,t,m_t^{2})\overline{v}(p_2)\Gamma_{i}^{\mu \nu} u(p_1) \epsilon_{3,\mu} \epsilon_{4,\nu},
\end{equation}
where $\mathcal{F}_{i}$ are the scalar form factors and $\Gamma_{i}^{\mu \nu}$ are the independent Lorentz tensors, which are built using only external momenta and polarisation vectors.

\noindent Since we are interested in the massive corrections that appears starting from the two-loop order, we are just interested in $\mathcal{F}_{i}^{(2)}$, i.e. the term $\mathcal{O}(\alpha_{s}^{2})$.

\begin{figure}[h!]
\includegraphics[width=\textwidth]{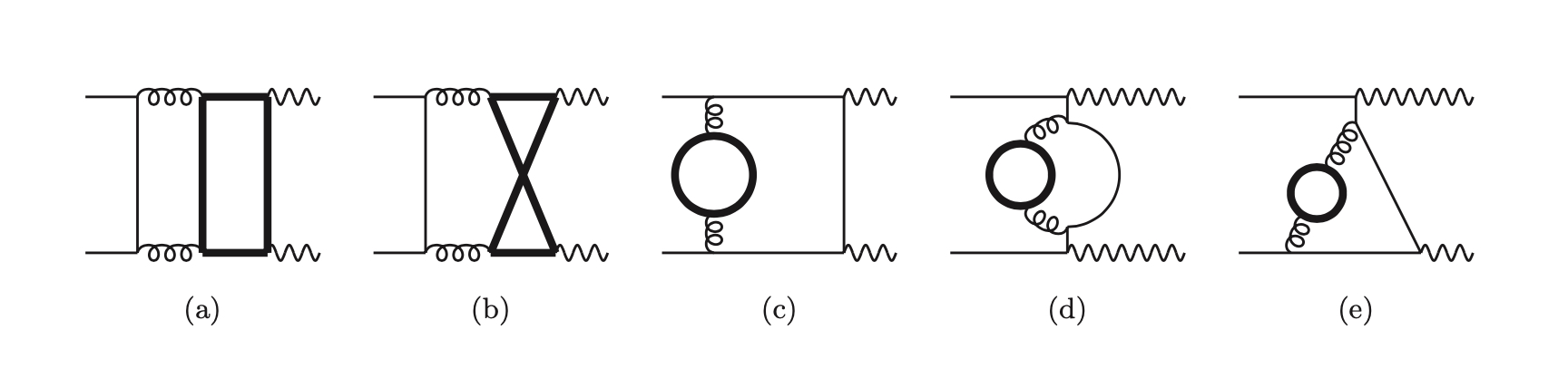}
  \caption{Representative set of two-loop diagrams with internal heavy-quark loops, which contribute at NNLO QCD  to diphoton production in the quark annihilation channel. Thin black lines represents light quarks, thick black lines heavy quarks, curly lines gluons and wavey lines photons.}
  \label{fig:diagrams}
\end{figure}

\noindent We decompose the form factors in terms of a basis of MIs, exploiting IBPs reduction.
 We generate the Feynman diagrams with FeynArts \cite{Hahn:2000kx} and  in Fig.~\ref{fig:diagrams}, we show a representative subset of them.

\noindent The MIs contributing to the process can be described by three different scalar integral topologies (modulo exchange of the two photons).  We have two planar topologies, PLA and PLB, and a non-planar topology NPL.
The PLA and PLB MIs were already studied in \cite{Becchetti:2017abb}, and three MIs of the NPL topology were already studied in \cite{vonManteuffel:2017hms}. Here we present as original result by itself a set of 9 new MIs of the NPL topology. This is exhaustive to describe all the processes  in terms of a minimal set of 42 MIs.

\section{Evaluation of the MIs}
\noindent The MIs are computed through the differential equations method. For the PLA family we put the system of differential equations in canonical logarithmic form \cite{Henn:2014qga} with respect to the kinematic invariants $\underline{x}=\{y,z\}$, where $y=\frac{s}{m_{t}^{2}}$, $z=\frac{t}{m_{t}^{2}}$.
The boundary conditions are given by $\underline{x}=0$, where all the MIs vanish except for two MIs, for which we use their analytical expressions.

\noindent The alphabet contains five different square roots of the kinematic invariants $r_{1},...,r_{5}$.
The canonical logarithmic form with the alphabet would allow us to obtain a fully analytic representation for the system of MIs. However, there are different problems, first of all these square roots are non simultaneously rationalizable, then the solutions is non trivial, furthermore it's possible to obtain big expressions that are not useful for our phenomenological purpose. 

\noindent The PLB family contains only one different MIs from the other two topologies, which we computed analytically. 

\noindent For the NPL family the situation is even more complicated. We write the system of differential
equations , that can be divided in two different subsets. 

\noindent The first one is given by the MIs whose differential equations can be putted in canonical logarithmic form, while the second subset contains the MIs whose analytic solution involve elliptic functions.
The elliptic subset contains two sectors, the first one is a non-planar triangle with a massive loop and the second sector is the top sectors of the NPL topology.

\noindent We have all the problems explained for PLA topology, with a set of nine different square roots of the kinematic invariants, $q_{1},...,q_{9}$. Furthermore, the integrals involve $elliptic$ $multiple$ $polylogarithms$ (eMPLs) kernels \cite{Broedel:2019hyg,Broedel:2017kkb,brown2013multiple}. 
The inhomogeneous part of the differential equations contain the elliptic triangle integrals of the first sector. Moreover, also the homogeneous part of the differential equations contains elliptic functions. This can be verified by studying the maximal cut of the double-box integral. We obtain a one-fold integral, where the elliptic curve is given by the fourth-degree polynomial in the integration variable $z_{8}$:
\begin{equation}
y_{c}^{2} = (z_{8}+t)(z_{8}+s+t)(z_{8}-z_{+})(z_{8}-z_{-}),
\end{equation}
where $z_{\pm} = \frac{1}{2}\left(-s - 2t \pm \sqrt{s}\sqrt{s + 16m_t^2}\right)$.
We observe that the elliptic curve obtained for the double-box degenerates to the same curve of the massive triangle in the forward limit $t=0$.

\noindent Since we are interested in phenomenological studies, to avoid all the problems coming from the analytical solution, we decided to solve semi-analitically the MIs through the generalised power series approach \cite{Moriello:2019yhu}. The numerical evaluation of the MIs has been made with DiffExp \cite{Hidding:2020ytt} and several check were made using the Mathematica package AMFlow \cite{Liu_2023}, obtaining a correspondance up to 200 digits .

\begin{figure}[h!]
\begin{center}
\includegraphics[width= 0.5 \textwidth]{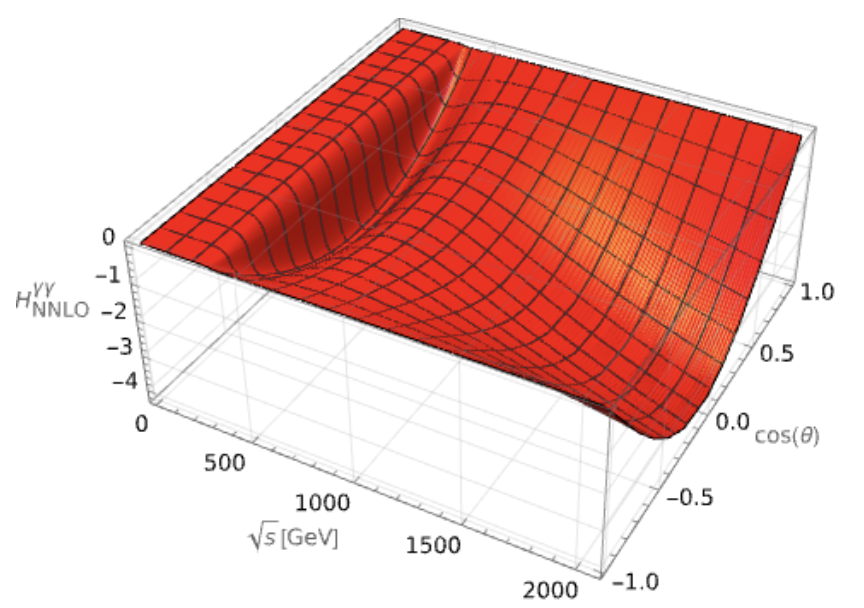}
\end{center}
  \caption{Numerical grid for $\mathcal{H}^{\gamma\gamma}_{\rm NNLO}$ as function of $\sqrt{s}$ and $cos(\theta)$}
  \label{fig:typesdiagrams}
\end{figure}

\noindent $\mathcal{F}_{i}^{(2)}$ does not have infrared (IR) poles, then after remove the ultraviolets (UV) poles, we can compute the NNLO Hard Function $ \mathcal{H}_{NNLO}^{\gamma\gamma}$ in the $q_{T}$-subtraction scheme \cite{Catani:2013tia}.
We proceed constructing a grid for the hard function covering the $2 \to 2$ physical space:
\begin{equation}
    s > 0, \;\;\; t=-\frac{s}{2}(1-\cos(\theta)), \;\;\; -s < t <0,
\end{equation}
where $\theta$, $0 < \theta < \pi$, is the scattering angle in the partonic center of mass frame.
The grid in Fig.~\ref{fig:typesdiagrams} was prepared for a total of 13752 points, with 24 different values of $cos(\theta)$ and 573 values of $\sqrt{s}$ in the range:
\begin{equation}
    -0.99 < \cos{\theta} < 0.99, \;\;\; 8 \operatorname{GeV} < \sqrt{s} < 2.2 \operatorname{TeV}.
\end{equation}
The evaluation time for PLA MIs is of $\mathcal{O}(2.5h)$ and $\mathcal{O}(10.5h)$ for NPL MIs, both on a single core.

\section{Full top-quark mass dependence in diphoton production at NNLO in QCD}

\noindent In this section we include the Double-Virtual contribution discussed in the previous sections with the Double-Real and Real-Virtual contributions in order to present our results for the diphoton production at NNLO in perturbative QCD taking into account the full top-quark mass dependence. 
We fix te value of the top quark mass at $m_{t}=173$ $GeV$.

\noindent We consider isolated diphoton production in $pp$ collisions at the centre-of-mass energy $\sqrt{s}=13$ $TeV$. We apply the fiducial cuts used in the last ATLAS collaboration \cite{ATLAS:2021mbt} and we use the smooth cone isolation criterion \cite{Frixione:1998jh,Frixione:1999gr}, in order to eliminates the entire fragmentation contribution.

\begin{figure}[h!]
\centering

   \includegraphics[width=0.46\textwidth]{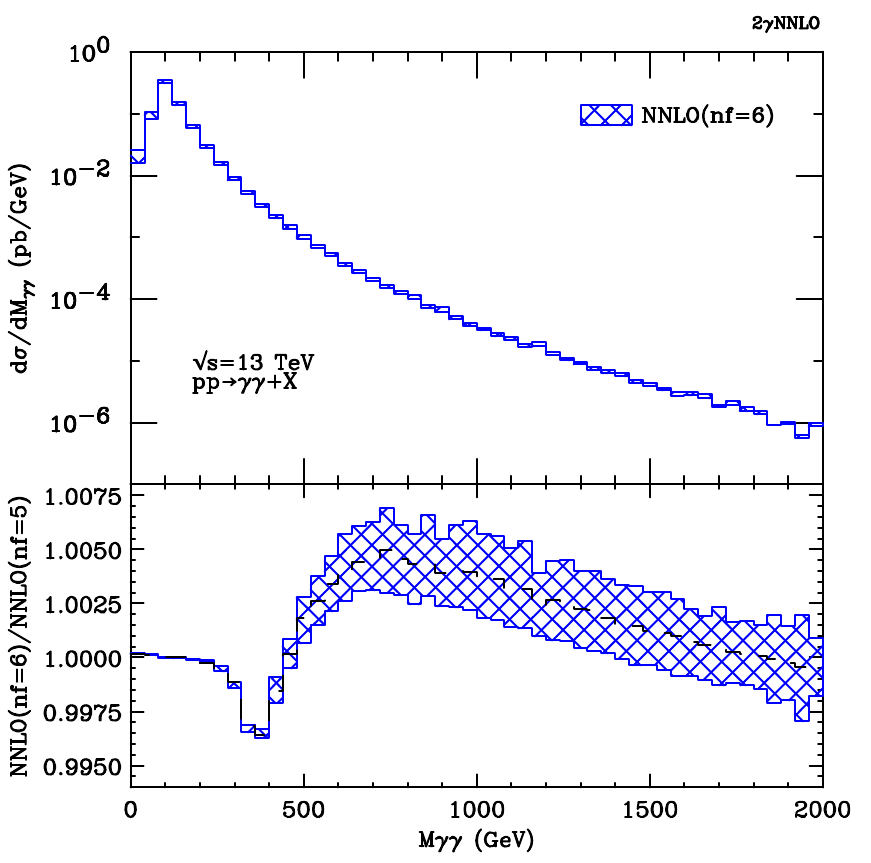}
\hfill
    \includegraphics[width=0.465\textwidth]{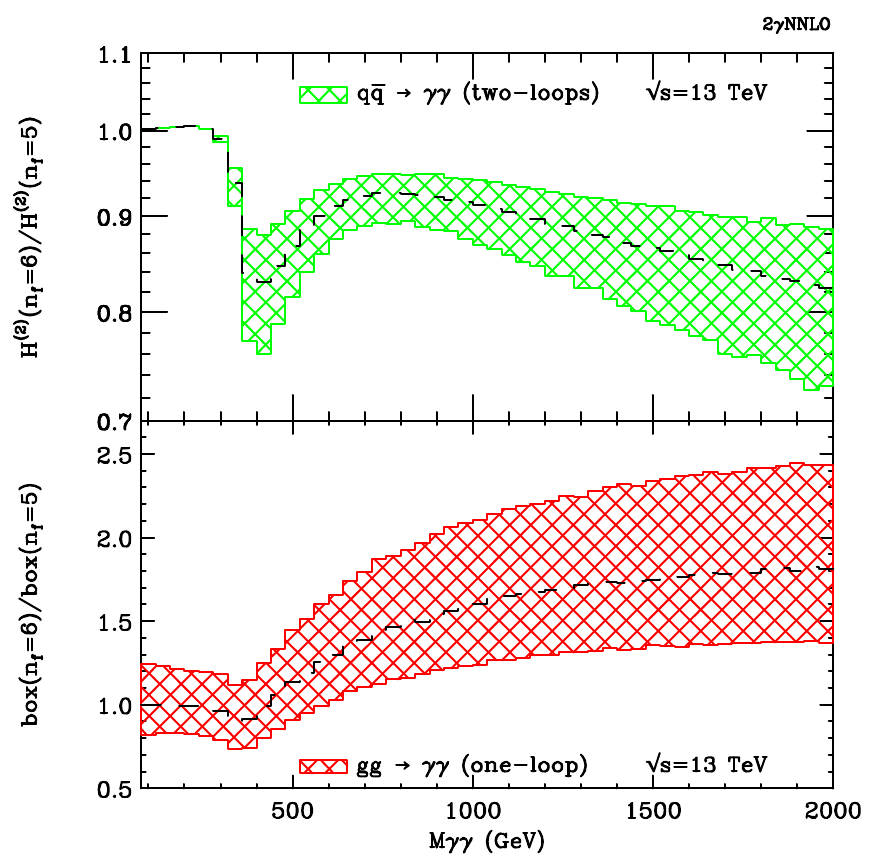}

\caption{NNLO invariant mass distribution with full top quark mass dependence on the left and ratio of the two-loop ($q\bar{q} \rightarrow \gamma\gamma$) massive form factor to the massless one with comparison of the 1-loop box ($gg\rightarrow \gamma\gamma$) contribution in the upper panel on the right. The central scale is shown with a black dashed line.}
\label{fig:final}
\end{figure}

\noindent Firstly, we comment the main contribution of our work, i.e. the Double-Virtual contribution in the invariant mass distribution. In Fig.~\ref{fig:final} (upper panel on the right) we show the ratio between the fully massive two-loop form factors and the already known massless case. The bands are computed implementing the 7-point scale variation. As expected, the ratio exhibits the typical peak around the top-quark threshold. In the lower panel we see the ration between the invariant mass distribution of massive box and the massless one. We observe that the ratios exhibits different shapes at large invariant mass.

\begin{figure}[h!] 
\centering

     \includegraphics[width=6cm]{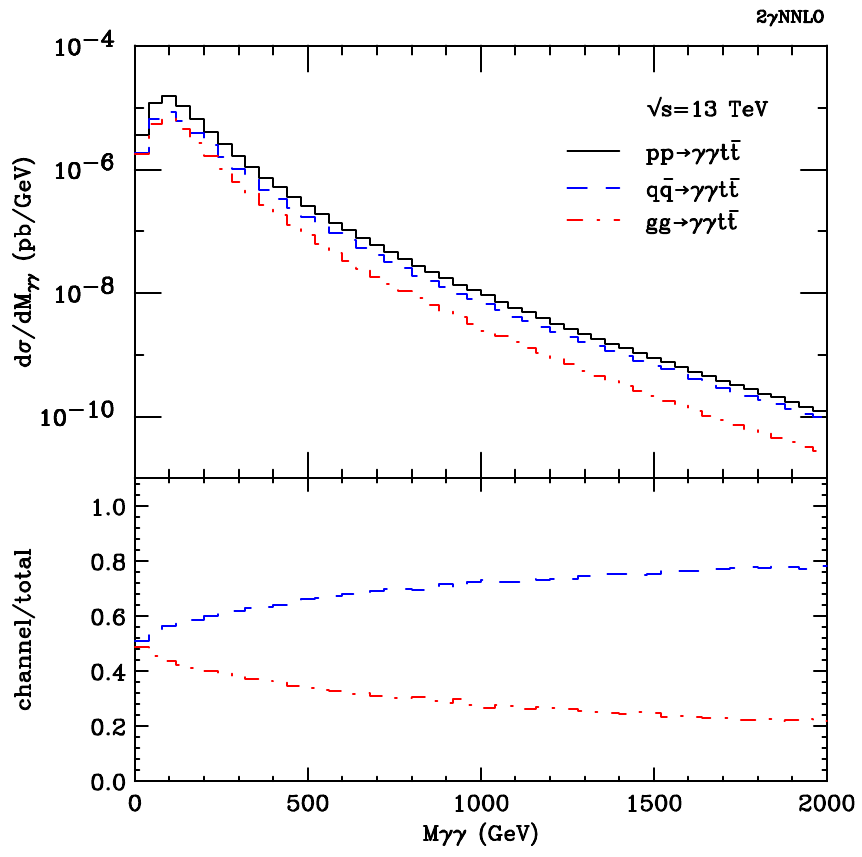}
\hfill
     \includegraphics[width=9cm]{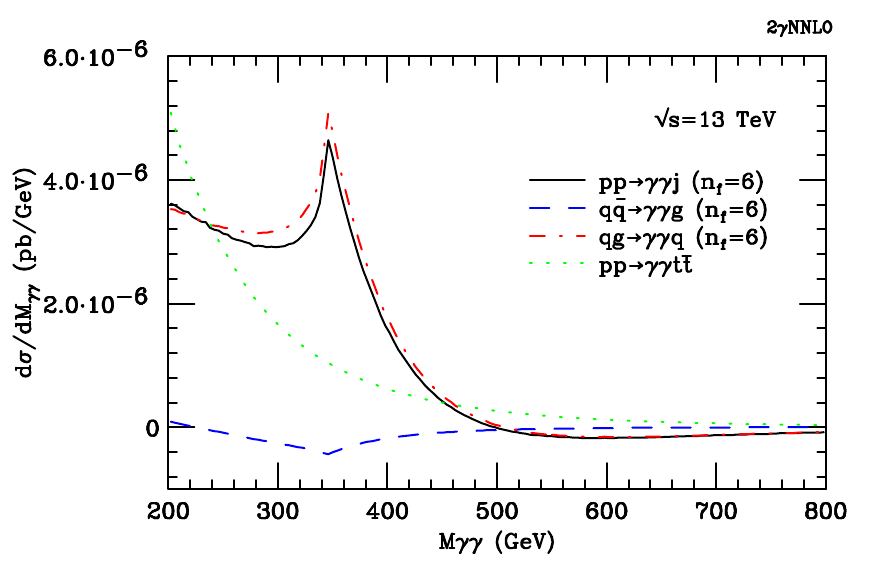}

\caption{
\label{fig:H2andBox}
{\em Invariant mass distribution for the Double-Real contributions on the left and the comparison with the Real-Virtual contributions on the right.} }
\label{fig:checks}
\end{figure}

\noindent We briefly discuss the massive Double-Real ($pp \to \gamma \gamma t \overline{t}$) and the Real-Virtual ($q \overline{q} \to \gamma \gamma g$ and $ q g \to \gamma \gamma q$) contributions. In Fig.~\ref{fig:H2andBox} we show on the left the invariant mass distribution obtained for the partonic sub-process. Since we produce two on-shell top-quarks and since we are computing tree-level scattering amplitudes, there is no top-quark threshold in the distribution. The only peak in the invariant mass distribution is due to kinematic effects, with the peaks at $2 p_{T \gamma}^{hard \hspace{0.1cm} cut}=80$ $GeV$, as in the massless case  \cite{Catani:2018krb}. In the bottom panel we show the relative size of the two different channels ($q\overline{q}$ and $gg$) with respect to the total. 

\noindent On the right we discuss the contribution of the one-loop NNLO massive contributions ($pp \to \gamma \gamma j$). We compare the invariant mass distributions of different channels with respect to the total correction. The $q \overline{q}$ and $qg$ channels show very different behaviour. The positive peak around the threshold is also found in the box contribution when only a massive quark is running in the loop. But in the box case the light quarks are also included in the loop, then the destructive interference between the different terms dominates the box contribution, producing the negative peak. In the massive Real-Virtual we consider the interference with the tree-level matrix elements, with only the massive top quark running in the loop, then there is no such mixing. 

\noindent In Fig.~\ref{fig:final}  (left panel) we present our final and complete results regarding the invariant mass distribution of diphoton production at NNLO in QCD using the fiducial cuts discussed above. We show the ratio between the fully massive NNLO result and the NNLO prediction for five light quark flavors. Around the top-quark threshold there is the negative peak due to a superposition of the effects from the different loop contributions. Beyond the negative peak, the massive NNLO prediction presents a positive deviation from the massless result at about 2.3 times of the value of the top-quark threshold. The effect of the massive corrections, with the fiducial cuts discussed above, in the range of invariant mass from 8~GeV to 2~TeV, is a deviation from the massless result in the range [-0.4\%, 0.8\%]. The effect can be bigger if we use different cuts and we extend the invariant mass range.

\section{Conclusions}
\noindent We presented the computation of two-loop massive form factors, the only missing ingredient for a full NNLO study. The MIs were evaluated through the generalised power series expansion, with a discussion about the elliptic behavior of the NPL topology MIs. We presented for the first time the complete NNLO QCD diphoton production taking into account the full top quark mass dependence, showing a detailed study of the impact of the massive corrections in the invariant mass distribution around the top quark threshold.

\section*{Acknowledgements}
\noindent This work is supported by the Spanish Government (Agencia Estatal de Investigaci\'on MCIN/AEI/ 10.13039/501100011033) Grant No. PID2020-114473GB-I00, and Generalitat Valenciana Grants No. PROMETEO/2021/071 and ASFAE/2022/009 (Planes Complementarios de I+D+i, Next Generation EU). 
F.C. is supported by Generalitat Valenciana GenT Excellence Programme (CIDE\-GENT/2020/011) and ILINK22045.

\newpage

\providecommand{\href}[2]{#2}\begingroup\raggedright\endgroup

\end{document}